\documentclass[aps,prl,twocolumn]{revtex4}
\usepackage{amsmath}
\usepackage{graphicx}
\begin{document}

\title{Intrinsic Coupling between Current and Domain Wall Motion in (Ga,Mn)As}

\author{Kjetil Magne D{\o}rheim Hals, Anh Kiet Nguyen, and Arne Brataas}

\affiliation{Department of Physics, Norwegian University of Science and 
Technology, NO-7491, Trondheim, Norway}

\begin{abstract}
  We consider current-induced domain wall motion and, the reciprocal process, moving domain wall-induced current. The associated Onsager coefficients are expressed in terms of scattering matrices.  Uncommonly, in (Ga,Mn)As, the effective Gilbert damping coefficient $\alpha_w$ and the effective out-of-plane spin transfer torque parameter $\beta_w$ are dominated by spin-orbit interaction in combination with scattering off the domain wall, and not scattering off extrinsic impurities. Numerical calculations give $\alpha_w \!  \sim \! 0.01$  and  $\beta_w \! \sim \! 1$ in dirty (Ga,Mn)As.   The extraordinary large $\beta_w$ parameter allows experimental detection of current or voltage induced by domain wall motion in (Ga,Mn)As. 
\end{abstract}

\maketitle

\newcommand{\eq}  {  \! = \!  }
\newcommand{\keq} {\!\! = \!\!}
\newcommand{\kadd}{  \! + \!  }
\newcommand{\ksim}{\! \sim \!}


The principle of giant magneto resistance is used to detect magnetic information. Large currents in magnetic nanostructures can manipulate the magnetization via spin transfer torques ~\cite{SpinTransfer}.  A deeper knowledge of the coupled out-of-equilibrium quasi-particle and magnetization dynamics is needed to precisely control and utilize current-induced spin transfer torques. 

The magnetization relaxes towards its equilibrium configuration by releasing magnetic moments and energy into reservoirs. This friction process is usually described by the Gilbert damping constant $\alpha$ in the Landau-Lifshitz-Gilbert (LLG) equation. Spins traversing a magnetic domain wall exert an in-plane and an out-of-plane torque on the wall~\cite{InOutPlaneTorque}. In \textit{dirty} systems, when the domain wall is wider than the mean-free-path, the out-of-plane torque, often denoted the non-adiabatic torque, is 
parameterized by the so-called $\beta$-factor~\cite{InOutPlaneTorque}. The Gilbert damping coefficient $\alpha$, the in-plane spin-transfer torque, and the out-of-plane torque coefficient $\beta$ determine how the magnetization is influenced by an applied current, e.g. the current-induced Walker domain wall drift velocity is proportional to $\beta / \alpha$~\cite{InOutPlaneTorque,CurrentIndDWMotion_eksp,Tserkovnyak:review07}. Scattering off impurities are important for $\alpha$ and $\beta$ ~\cite{InOutPlaneTorque,CurrentIndDWMotion_eksp,Tserkovnyak:review07}. Additionally, domain wall scattering can contribute to $\alpha$ and $\beta$. In \textit{ballistic} (Ga,Mn)As, intrinsic spin-orbit coupling causes significant hole reflection at the domain wall, even in the adiabatic limit when the wall is much thicker than the Fermi wavelength~\cite{Nguyen:prl06}. This grossly increases the out-of-plane spin-transfer torque, and consequently the current-driven domain wall mobility.  So far, there are no investigations on the effect of these domain wall induced hole reflections on the effective Gilbert damping constant $\alpha$.

Experimental (Ga,Mn)As samples are dirty so that the effect of disorder on the effective Gilbert damping and the out-of-plane spin transfer torque should be taken into account. We find surprisingly that, in systems with a large intrinsic spin-orbit coupling, domain wall scattering contributes dominantly to $\alpha$ and $\beta$ even in the dirty limit. Intrinsic current-domain wall motion coupling is robust against impurity scattering.

Current-induced domain-wall motion has been seen in many experiments~\cite{CurrentIndDWMotion_eksp}. The reciprocal effect, domain-wall motion induced current, is currently theoretically investigated~\cite{PrecessingWallInducedCurrent,Duine:prb08}, and seen experimentally~\cite{Yang:prl09}. A precessing domain wall induces a charge current in ferromagnetic metals~\cite{PrecessingWallInducedCurrent} similar to spin-pumping in layered ferromagnet-normal metal systems~\cite{GilbertDampingAndSpinPumping}. For rigid domain wall motion, the induced charge current is proportional to $\beta / \alpha$~\cite{Duine:prb08}. We find that $\beta$ and $\beta/\alpha$ in (Ga,Mn)As are so large that the current, or equivalently, the voltage induced by a moving domain wall is experimentally measurable. 

Onsager's reciprocity relations dictate that response coefficients of domain wall motion induced current and current induced domain wall motion are related. In dirty systems, these relations have been discussed in Ref.~\cite{Duine:prb08}. Ref.~\cite{Duine:prb08} also used the scattering theory of adiabatic pumping to evaluate the non-adiabatic spin-transfer torque in ballistic systems without intrinsic spin-orbit interaction. We first extend the pumping approach to (Ga,Mn)As with strong intrinsic spin-orbit interaction, and second, also evaluate the Onsager coefficient as a function of sample disorder. In determining all Onsager coefficients, magnetization friction must be evaluated on the same footing. To this end, we generalize the energy pumping scattering theory of Gilbert damping \cite{Brataas:prl08} to domain wall motion. Our numerical calculation demonstrates, for the first time, that domain wall scattering is typically more important than impurity scattering for the effective domain wall motion friction in systems with a strong intrinsic spin-orbit interaction. Our novel Onsager scattering approach can also be used to compute the effective rigid domain wall motion $\alpha$ and $\beta$ parameters in realistic materials like Fe, Ni, Co, and alloys thereof from first-principles.

Let us discuss in more detail the Onsager reciprocity relations in our system. The magnetic field is a thermodynamic force for the magnetization since it can move domain walls. The electric field is a thermodynamic force for the charges as it induces currents. In systems where charge carriers also carry spin, the magnetic and charge systems are coupled. Through this coupling, the electric field can move a domain wall and, vice versa, the magnetic field can induce a current. This phenomenon, where the thermodynamic force of one system can induce a flux in another system is well-known in thermodynamics~\cite{de_Groot:bok}: Assume a  system described by the quantities $\left\{ q_i \right\}$, $X_i$ denotes the thermodynamic force, and $J_i$ the flux associated with the quantity $q_i$.  In linear response, $J_i= \sum_j L_{ij}X_j$, where $L_{ij}$ are the Onsager coefficients.  Onsager's reciprocity principle dictates $L_{ij}= \epsilon_i\epsilon_j L_{ji}$, where $\epsilon_i=1$ ($\epsilon_i=-1$) if $q_i$ is even (odd) under time-reversal~\cite{de_Groot:bok}. Fluxes and forces are not uniquely defined, but the Onsager reciprocity relations are valid when the entropy generation is $\dot{S}= \sum_i J_i X_i$~\cite{de_Groot:bok}. 

We first derive expressions for the Onsager coefficients and determine the Onsager reciprocity relations between a charge current and a moving domain wall in terms of the scattering matrix. Subsequently, we derive the relation between the Onsager coefficients and the effective Gilbert damping parameter $\alpha_w$ and the out-of-plane torque parameter $\beta_w$  for domain wall motion. Finally, we numerically compute $\alpha_w$ and $\beta_w$ for (Ga,Mn)As. 

We start the derivation of the Onsager coefficients in terms of the scattering matrix by assuming the following free energy functional for the magnetic system
\begin{eqnarray}
  F[\mathbf{M}]=M_s\int \!\! d\mathbf{r}\biggl( \frac{J}{2}[(\nabla\theta)^2 +
  \sin^2(\theta)(\nabla\phi)^2] + 
  \nonumber \\
  \frac{K_{\bot}}{2}\sin^2(\theta)\sin^2(\phi) - \frac{K_z}{2}\cos^2(\theta) -
  H_{\text{ext}}\cos(\theta) \, \biggr) ,
\label{MagnEnergy}
\end{eqnarray}
where $M_s$, $J$ and $H_{\text{ext}}$ are the saturation magnetization, spin-stiffness and external magnetic field, respectively, and $K_z$ and $K_{\bot}$ are magnetic anisotropy constants. The local magnetization angles $\theta$ and $\phi$ are defined with respect to the $z$- and $x$-axis, respectively. The system contains a Bloch wall rotating in the (transverse) $x$-$z$ plane, $\cos(\theta)= \tanh([y-r_{w}]/\lambda_w)$, $\sin(\theta)= 1/\cosh([y-r_{w}]/\lambda_w)$,  where $r_w$ is the  position of the wall, and $\lambda_w$ is the wall width. We assume the
external magnetic field is lower than the Walker threshold, so that the wall rigidly moves ($\dot{\phi}=0$) with a constant drift velocity. In this case $r_w$ and $\phi$ completely characterize the magnetic system, and $\lambda_w= \sqrt{J/(K_z + K_{\bot}\sin^2(\phi))}$~\cite{Tserkovnyak:review07}. The current is along the $y$-axis.

The heat dissipated per unit time from a charge current $J_c$ is $\dot{Q}=J_c(V_L - V_R)$, where $V_L$ ($V_R$) is the voltage in the left (right) reservoir. Using the relation $dS=dQ/T$, this implies an entropy generation $\dot{S}= J_c(V_L - V_R)/T$.  Thus, $X_c\equiv(V_L - V_R)/T$ is the thermodynamic force inducing the flux $J_c$. We assume the magnetic system to be at constant temperature, which means that the heat transported out of the magnetic system as the domain wall moves equals the loss of free energy. This implies an entropy generation $\dot{S}= \dot{Q}/T= -\dot{F}/T= (-\partial F[r_w,\phi]/T\partial r_w)\,\dot{r}_w= X_w J_w$, where we have defined the force $X_w\equiv -\partial F[r_w,\phi]/T\partial r_w$ and flux $J_w\equiv\dot{r}_{w}$. Using Eq.~\eqref{MagnEnergy}, we find $X_w= -2 A M_sH_{\text{ext}}/T$, where $A$ is the conductor's cross-section. Fluxes are related to forces by 
\begin{eqnarray}
J_w & = & L_{ww}  X_w + L_{wc} X_c \label{eqn:J_w}\\
J_c & = & L_{cc}X_c + L_{cw} X_w, \label{eqn:J_N}
\end{eqnarray} 
where $L_{cc} = G T$ and $G$ is the conductance. $L_{ww}$ ($L_{wc}$) determine the induced domain wall velocity by an external magnetic field (a current). The induced current by a moving domain wall caused by an external magnetic field $H_{\text{ext}}$ is controlled by $L_{cw}$. Both charge and $r_w$ are even under time-reversal so that $L_{cw}=L_{wc}$~\cite{comment1}.

The current induced by a moving domain wall is parametric pumping in terms of the scattering matrix~\cite{GilbertDampingAndSpinPumping}:
\begin{equation}
  J_{c,\!\alpha} = \frac{e\dot{r}_w}{2\pi}\sum_{\beta=1,2}\Im m\left\{ \text{Tr}\left[ 
       \frac{\partial S_{\alpha\beta}}{\partial r_w}  S_{\alpha\beta}^{\dagger}  \right]   \right\} \, ,
\label{eq:Brouwer}
\end{equation}
where $S_{\alpha\beta}$ is the scattering matrix between transverse modes in lead $\beta$ to transverse modes in lead $\alpha$. The system  has two leads ($\alpha , \beta \in \left\{ 1,2 \right\} $). The trace is over all propagating modes at the Fermi energy $E_F$.   From Eqs.~\eqref{eqn:J_w} and \eqref{eqn:J_N} we find $J_c= L_{cw}\dot{r}_w/L_{ww}$.

We consider transport well below the critical transition temperature in (Ga,Mn)As, which is relatively low, and assume the energy loss in the magnetic system is transferred into the leads by holes. Generalizing Ref.\ \cite{Brataas:prl08} to domain wall motion, this energy-flux is related to the scattering matrix:
\begin{eqnarray}
  J_E= \frac{\hbar}{4\pi} \text{Tr}\left\{ \frac{dS}{dt}\frac{dS^{\dagger}}
    {dt}   \right\} 
  = \frac{\hbar\dot{r}_w^2}{4\pi} \text{Tr}\left\{ \frac{\partial S}{\partial r_w}
    \frac{\partial S^{\dagger}}{\partial r_w}   \right\}.
\end{eqnarray}
For a domain wall moved by an external magnetic field, we then find
that $X_wJ_w= J_w^2/L_{ww}=J_E/T$. In summary, the Onsager coefficients in Eq.~\eqref{eqn:J_w} and Eq.~\eqref{eqn:J_N} are
\begin{eqnarray}
  L_{ww} & = &  \left(  \frac{\hbar}{4\pi } \text{Tr}\left\{ 
      \frac{\partial S}{\partial r_w}\frac{\partial S^{\dagger}}{\partial r_w}   \right\}  \right)^{-1} 
  \label{Eq:L_rr} \, , \\
  L_{cw} &=&  \frac{2e}{\hbar} \frac{\sum_{\beta=1,2}\Im m\left\{ \text{Tr}\left[ 
        \frac{\partial S_{\alpha\beta}}{\partial r_w}  S_{\alpha\beta}^{\dagger}    \right]  
         \right\}}{\text{Tr} \left\{ \frac{\partial S}{\partial r_w}
      \frac{\partial S^{\dagger}}{\partial r_w}   \right\}} \, , \label{Eq:L_rN}  \\
  L_{cc} & = & \frac{e^2}{h} \text{Tr } \left\{t^{\dagger}t  \right\} \, , \label{Eq:L_cc}
\end{eqnarray} 
where $t$ is the transmission coefficient in the scattering matrix. We have omitted the temperature factor in the coefficients (\ref{Eq:L_rr}), (\ref{Eq:L_rN}), and (\ref{Eq:L_cc}) since it cancels with the temperature factor in the forces, \textit{i.e.} we transform $L \rightarrow L/T$ and $X\rightarrow T X$. 
The Onsager coefficient expressions in terms of the scattering matrix are valid irrespective of impurity disorder and spin-orbit interaction in the band structures or during scattering events, and can treat transport both in ballistic and diffusive regimes.

Let us compare the global Onsager cofficients (\ref{Eq:L_rr}), (\ref{Eq:L_rN}), and (\ref{Eq:L_cc}) with the local Onsager coefficients in the dirty limit to gain additional understanding. In the dirty limit, all Onsager cofficients become local and the magnetization dynamics can be described by the following phenomenological local LLG equation~\cite{InOutPlaneTorque,CurrentIndDWMotion_eksp}:
\begin{eqnarray}
  \dot{\mathbf{m}} &=& -\gamma\mathbf{m}\times
  \mathbf{H}_{\text{eff}} + \alpha \mathbf{m} \times \dot{\mathbf{m}}  \nonumber \\ 
 & & - \left( 1-  \beta\mathbf{m}\times   
  \right)\left(\mathbf{v}_s \cdot \nabla \right) \mathbf{m},	\label{LLG}
\end{eqnarray}
where  $\mathbf{m}$ is the magnetization direction, $\mathbf{H}_{\text{eff}}$ is the effective magnetic field, $\gamma$ is the gyromagnetic ratio, $\mathbf{v}_s= -\hbar P \mathbf{j} /(e S_0)$, $S_0= M_s/\gamma$, $M_s$ the magnetization, $\alpha$  the Gilbert damping constant, $P$ the  spin-polarization along $-\mathbf{m}$ of the charge carriers~\cite{comment2}, and $\beta$ is the out-of-plane spin-transfer torque parameter.  Substituting a Walker ansatz into Eq.~\eqref{LLG} gives below the Walker threshold ~\cite{Tserkovnyak:review07}: $\alpha\dot{r}_{w}/ \lambda_w = -\gamma H_{\text{ext}} - \hbar \beta P j / \left( eS_0 \lambda_w\right)$.	
In dirty, local, systems this equation determines the relation between the flux $J_w$ and the forces $X_w$ and $X_c$ as $L_{ww}= \lambda_w/(2AS_0\alpha)$ and $L_{wc}=-\hbar \beta P G/(e\alpha S_0 A)$, where we have used $j=\sigma (V_L - V_R)/L$, and $G=\sigma
A/L$. Here, $L$ is the length of the conductor, $e$ the electron charge, and $\sigma$ the conductivity. This motivates defining the following dimensionless global coefficients:
\begin{eqnarray}
  \alpha_{w} \equiv \frac{\lambda_w}{2AS_0L_{ww}}, & &
  \beta_{w} \equiv -\frac{\lambda_w e}{2\hbar P G}\frac{L_{wc}}{L_{ww}}. 
\nonumber
\end{eqnarray}
$\alpha_w$ is the effective Gilbert damping coefficient and $\beta_w$ is the effective out-of-plane torque on the domain wall.

We will in the following investigate $\alpha_w$ and $\beta_w$ for (Ga,Mn)As by calculating the scattering matrix expressions in Eq.~\eqref{Eq:L_rr} and Eq.~\eqref{Eq:L_rN}.  We use the following Hamiltonian to model quantum transport of itinerant holes:
\begin{equation}
  H = H_{\text{L}}
  + \mathbf{h(r)}\cdot \mathbf{J} + V(\mathbf{r}) \, .
\label{Hamiltonian}
\end{equation}
Here, $H_{\text{L}}$ is the $4 \!  \times \!  4$ Luttinger Hamiltonian (parameterized by $\gamma_1$
and $\gamma_2$) for zincblende semiconductors in the spherical approximation, while  $\mathbf{h}\cdot\mathbf{J}$  describes the exchange interaction between the itinerant holes and the local magnetic moment of the Mn dopants. We introduce Anderson impurities as $V(\mathbf{r})= \sum_i V_i\delta_{\mathbf{r},\mathbf{R_i}}$, where $\mathbf{R_i}$ is the position of impurity $\,i$, $V_i$ its impurity strength, and $\delta$ the Kronecker delta. More details about the model and the numerical method used can be found in Refs.~\cite{Jungwirth:RMP06, Nguyen:prl08}. 

We consider a discrete conductor with transverse dimensions $L_x=23nm$, $L_z=17nm$ and length $L_y=400nm$. The lattice constant is $1nm$, much less than the typical Fermi wavelength $\lambda_F \sim 8nm$. The Fermi energy $E_F=82meV$ is measured from the bottom of the lowest subband. $|\mathbf{h}|=0.5\times 10^{-20}J$ and $\gamma_1=7$. The typical mean-free path for the systems studied ranges from the diffusive to the ballistic regime $l \sim 23nm \rightarrow \infty$, and we are in the metallic regime $k_F l \gg 1$. The domain wall length is $\lambda_w=40nm$. The spin-density $S_0$ from the local magnetic moments is $S_0= 10 \hbar x / a^3_{GaAs}$, $a_{\text{GaAs}}$ the lattice constant for GaAs, and $x=0.05$ the doping level\cite{Jungwirth:RMP06}.

\begin{figure}[h]
\centering
\includegraphics[scale=1.0]{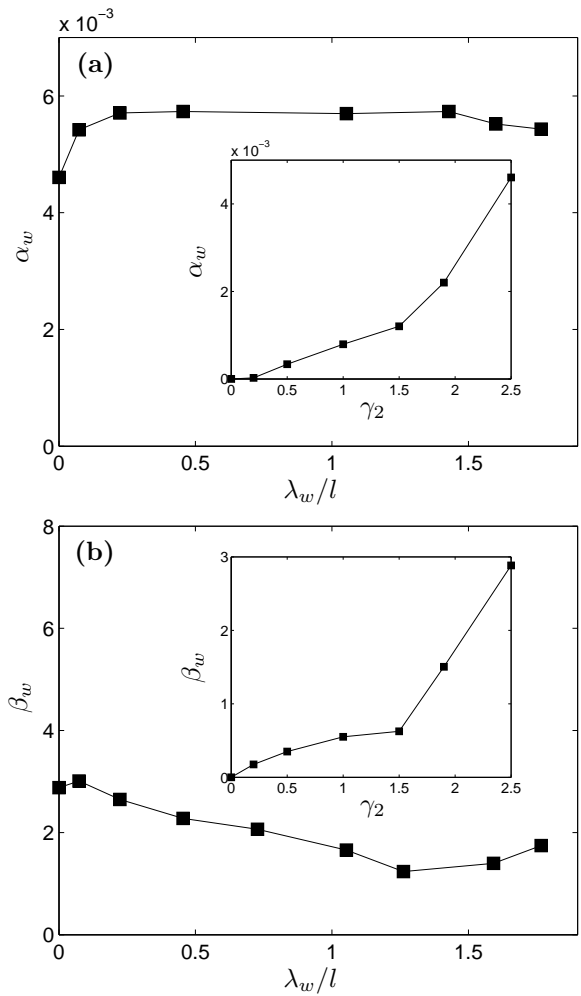}
\caption{\textbf{(a)}: Effective Gilbert damping $\alpha_w$ as function of $\lambda_w/l$,
    where $\lambda_w$ is the domain wall length and $l$ is the mean
    free path when $\gamma_2=2.5$. Here, $\lambda_w$ is kept fixed,
    and $l$ is varied.  Inset:
    $\alpha_w$ as a function of spin-orbit coupling
    $\gamma_2$ for a clean system, $l=\infty$.  
    \textbf{(b)}: $\beta_w $ as a function of
    $\lambda_w/l$, where $\lambda_w$ is the domain wall length and $l$
    is the mean free path when $\gamma_2=2.5$. Here, $\lambda_w$ is
    kept fixed, and $l$ is varied. Inset:
    $\beta_w$ as a function of spin-orbit
    coupling $\gamma_2$ for a clean system, $l=\infty$. In all plots, line is guide to the eye.  }
\label{fig:alfabeta}
\end{figure}
Fig.~\ref{fig:alfabeta}a shows the computed effective Gilbert damping coefficient $\alpha_w$
versus $\lambda_w/l$ for (Ga,Mn)As containing one Bloch wall. Note the relatively high $\alpha_w \sim 5 \times 10^{-3}$ in the ballistic limit. Additional impurities, in combination with the spin-orbit coupling, assist in releasing energy and angular momentum into the reservoirs and increase $\alpha_w$. However, as shown in Fig.~\ref{fig:alfabeta}a, impurities contribute only about $20\%$ to $\alpha_w$ even when the domain wall is two times longer than the mean free path. Due to the strong spin-orbit coupling, ballistic domain walls have a large intrinsic resistance~\cite{Nguyen:prl06} that survives the adiabatic limit. When itinerant holes scatter off the domain wall their momentum changes and through the spin-orbit coupling their spin also changes. This is the dominate process for releasing energy and magnetization into the reservoirs.  The saturated value $\alpha_w \sim 6\times 10^{-3}$ is of the same order as the estimates in Ref.~\cite{Garate:cm08} for bulk (Ga,Mn)As.  The inset in Fig.~\ref{fig:alfabeta}a shows the domain-wall contribution to $\alpha_w$ versus the spin-orbit coupling for a clean system with no impurities. $\alpha_w$ monotonically decreases for decreasing $\gamma_2$ and vanish for $\gamma_2 \rightarrow 0$. Since, $\lambda_w/\lambda_F \ksim 5$, itinerant holes will, without spin-orbit coupling, traverse the domain wall adiabatically. 

Fig. \ref{fig:alfabeta}b shows $\beta_w$ versus $\lambda_w/l$. $\beta_w$ decreases with increasing disorder strength. This somewhat counter intuitive result stem from the fact that domain walls in
systems with spin-orbit coupling have a large intrinsic domain wall resistance~\cite{Nguyen:prl06} which originates from the anisotropy in the distribution of conducting channels~\cite{Nguyen:prl06}. The reflected spins do not follow the magnetization of the domain wall, and thereby cause a large out-of-plane torque~\cite{InOutPlaneTorque}. This causes the large $\beta_w$ in the ballistic limit. Scalar, rotational symmetric impurities tend to reduce the anisotropy in the conducting channels, and thereby reduce the intrinsic domain wall resistance and consequently reduce $\beta_w$. Deeper into the diffusive regime, $\beta$ saturates. Here, the domain wall resistance and $\beta_w$ are kept at high levels due to the increase in the spin-flip rate caused by impurity scattering. The saturated value is $\beta \ksim 1$. For even dirtier systems than a reasonable computing time allows, we expect a further increase in $\beta_w$. In comparison, simple microscopic theories for ferromagnetic metals where one disregards the spin-orbit coupling in the band structure predict $\beta\sim 0.001-0.01$ ~\cite{InOutPlaneTorque,CurrentIndDWMotion_eksp,Tserkovnyak:review07}. Similar to the Gilbert damping, in ballistic systems $\beta_w$ increases with spin-orbit coupling because of the increased domain wall scattering~\cite{Nguyen:prl06}, see Fig.~\ref{fig:alfabeta}b inset.

$\beta_w$ can be measured experimentally by the induced current or voltage from a domain wall moved by an external magnetic field as a function of the domain wall velocity~\cite{Duine:prb08}. From the Onsager relations we have that $J_c= L_{cw}X_w$. Using $X_w= J_w/L_{ww}$, the induced current and voltage are~\cite{Duine:prb08}:
\begin{equation}
  J_c = - 2\beta\frac{\hbar P G}{e\lambda_w}\dot{r}_{w}\ \Rightarrow\  V = 
  - 2\beta_w \frac{\hbar P}{e\lambda_w}\dot{r}_{w} \label{CurrentFromOnsager} \, .
\end{equation}
An estimate of the maximum velocity of a domain wall moved by an external magnetic
field below the Walker treshold is $\dot{r}_w\sim 10~m/s$~\cite{Dourlat:prb08}. With $\lambda_w=40~nm$ and $P=0.66$ this indicates an experimentally measurable voltage $V\sim 0.2~\mu V$.

In conclusion, we have derived Onsager coefficients and reciprocity relations between current and domain wall motion in terms of scattering matrices. In (Ga,Mn)As, we find the effective Gilbert damping constant $\alpha_w \ksim 0.01$ and out-of-plane spin transfer torque
parameter $\beta_w \ksim 1$. In contrast to ferromagnetic metals, the main contributions to $\alpha_w$ and $\beta_w$ in (Ga,Mn)As are intrinsic, and induced by scattering off the domain wall, while impurity scattering is less important. The large $\beta_w$ parameter implies a measurable moving domain wall induced voltage.

This work was supported in part by the Research Council of Norway, Grants Nos.~158518/143 and 158547/431, computing time through the Notur project and EC Contract IST-033749 "DynaMax".


\begin{thebibliography}{99}
\bibitem{comment1}
The magnetic texture is inverted in this equation, i.e. $L_{cw}\left[ \mathbf{m} \right]= L_{wc} \left[ -\mathbf{m} \right]$.
\bibitem{comment2}
We define the spin-polarization $P= \left\langle J_z \right\rangle =\sum_j (\psi_j^{\dagger}J_z\psi_j)/N$ ($\mathbf{m}=-\mathbf{\hat{z}}$), where the sum is over $N$ propagating modes,  $\psi_j$ are the corresponding spinor-valued wavefunctions, and $J_z$ the dimensionless angular momentum operator.
\end{thebibliography}
\end{document}